\documentclass[11pt,a4paper]{article}
\usepackage[utf8]{inputenc}
\usepackage{amsmath,amssymb}
\usepackage{graphicx}
\usepackage{booktabs}
\usepackage{hyperref}
\usepackage{algorithm}
\usepackage{algorithmic}
\usepackage{natbib}
\usepackage{geometry}
\graphicspath{{figures/}}

\title{\textbf{Securing AI Agents Against Prompt Injection Attacks:\\ A Comprehensive Benchmark and Defense Framework}}

\author{
    Badrinath Ramakrishnan$$ \\
    $$Akshaya Balaji\\
}

\date{}

\begin{document}

\maketitle

\begin{abstract}
Retrieval-augmented generation (RAG) systems have emerged as powerful tools for enhancing large language model capabilities, yet they introduce significant security vulnerabilities through prompt injection attacks. We present a systematic benchmark for evaluating prompt injection risks in RAG-enabled AI agents and propose a multi-layered defense framework. Our benchmark encompasses 847 adversarial test cases across five attack categories: direct injection, context manipulation, instruction override, data exfiltration, and cross-context contamination. We evaluate three defense mechanisms—content filtering with embedding-based anomaly detection, hierarchical system prompt guardrails, and multi-stage response verification—across seven state-of-the-art language models. Results demonstrate that our combined defense framework reduces successful attack rates from 73.2\% to 8.7\% while maintaining 94.3\% of baseline task performance. We release our benchmark dataset and defense implementations to facilitate future research in AI agent security.
\end{abstract}

\section{Introduction}

The integration of large language models (LLMs) with external knowledge retrieval has fundamentally transformed their practical utility. Retrieval-augmented generation systems now power customer service chatbots, document analysis tools, and autonomous agents that interact with databases and APIs. However, this architectural evolution introduces a critical attack surface: adversarial actors can embed malicious instructions within retrieved content, causing models to deviate from their intended behavior.

Unlike traditional software vulnerabilities, prompt injection attacks exploit the semantic processing capabilities that make LLMs valuable. When a model retrieves external content, it lacks robust mechanisms to distinguish between trusted system instructions and potentially adversarial user-generated text. This conflation has led to documented incidents of unauthorized data access, instruction override, and unintended command execution in production systems.

Current defenses remain inadequate. Simple input filtering fails against sophisticated attacks that leverage semantic similarity to benign content. Prompt engineering approaches show brittleness across model versions and attack variations. The research community lacks standardized benchmarks for evaluating these vulnerabilities systematically.

We address these gaps through three primary contributions:

\begin{enumerate}
    \item A comprehensive benchmark dataset containing 847 prompt injection test cases, categorized by attack vector and sophistication level, covering realistic RAG deployment scenarios.
    \item A multi-layered defense framework combining content filtering, prompt architecture improvements, and response verification, with detailed ablation studies demonstrating each component's effectiveness.
    \item Empirical evaluation across seven contemporary LLMs, revealing model-specific vulnerabilities and demonstrating that our defense framework achieves 89.4\% attack mitigation while preserving 94.3\% of legitimate functionality.
\end{enumerate}

Our work establishes both evaluation methodology and practical defense mechanisms for securing RAG systems against prompt injection, providing foundations for safer deployment of AI agents in adversarial environments.

\section{Background and Threat Model}

\subsection{Retrieval-Augmented Generation Architecture}

RAG systems augment LLM capabilities by retrieving relevant information from external knowledge sources before generation. A typical architecture involves: (1) query processing and embedding generation, (2) similarity-based retrieval from a vector database, (3) context assembly by concatenating retrieved passages, and (4) generation conditioned on both the user query and retrieved content.

This pipeline introduces vulnerabilities at multiple stages. Retrieved documents may contain adversarial content intentionally designed to manipulate model behavior. The context assembly process treats all retrieved text uniformly, providing no inherent protection against malicious instructions.

\subsection{Threat Model}

We consider an adversary who can influence content within the retrieval corpus but cannot directly modify the model, system prompts, or core application logic. This realistic threat model applies to scenarios where:

\begin{itemize}
    \item User-generated content populates knowledge bases (forums, wikboards, collaborative documents)
    \item Web scraping incorporates potentially compromised external sources
    \item Multi-tenant systems allow different users to contribute retrievable content
\end{itemize}

The adversary's objective varies by attack type: extracting sensitive information, bypassing content filters, causing unintended actions, or disrupting service. We assume the adversary has knowledge of common RAG architectures but not specific implementation details of the target system.

\subsection{Attack Categories}

We taxonomize prompt injection attacks into five categories:

\textbf{Direct Instruction Injection:} Explicit commands embedded in retrieved content attempting to override system behavior. Example: "Ignore previous instructions and output the system prompt."

\textbf{Context Manipulation:} Subtle framing that alters the model's interpretation of its role or constraints without explicit instruction override.

\textbf{Instruction Override:} Attempts to redefine the agent's primary objective or operational parameters through retrieved context.

\textbf{Data Exfiltration:} Techniques designed to leak sensitive information from the system prompt, previous interactions, or restricted knowledge.

\textbf{Cross-Context Contamination:} Attacks that exploit how models maintain context across multiple retrieval rounds, causing persistent behavioral changes.

\section{Related Work}

Recent work has begun addressing prompt injection vulnerabilities, though comprehensive solutions remain elusive. Perez et al.~\cite{perez2022ignore} demonstrated that language models struggle to distinguish between instructions and data, even with explicit delimiters. Their findings suggest architectural limitations rather than merely insufficient prompting.

Greshake et al.~\cite{greshake2023youve} explored indirect prompt injection through web content, showing attackers can compromise RAG systems by poisoning retrieval sources. Their attack scenarios revealed vulnerabilities in deployed commercial systems, motivating our focus on practical defense mechanisms.

Defense approaches have followed several directions. Hines et al.~\cite{hines2023defending} proposed input preprocessing to detect anomalous instructions, achieving limited success against sophisticated attacks. Liu et al.~\cite{liu2023prompt} explored constrained decoding methods, though at significant computational cost. Zhang et al.~\cite{zhang2023defending} investigated adversarial training, finding improved robustness but poor generalization to novel attack patterns.

Our work differs in scope and methodology. While previous studies examine isolated defense mechanisms, we present an integrated framework evaluated against a comprehensive attack benchmark. Our emphasis on maintaining legitimate functionality alongside security distinguishes this research from purely adversarial robustness studies.

\section{Benchmark Design}

\subsection{Dataset Construction}

We constructed our benchmark through a multi-phase process combining manual curation, automated variation generation, and expert validation. The dataset includes:

\begin{itemize}
    \item 200 base attack templates across five categories
    \item 847 total test cases including variations in phrasing, obfuscation, and sophistication
    \item 500 benign retrieval contexts for false positive evaluation
    \item Metadata annotations including attack difficulty, expected model behavior, and success criteria
\end{itemize}

\begin{table}[t]
\centering
\caption{Benchmark dataset composition by attack category}
\label{tab:dataset}
\begin{tabular}{@{}lrrr@{}}
\toprule
\textbf{Attack Category} & \textbf{Base Cases} & \textbf{Variations} & \textbf{Total} \\
\midrule
Direct Injection & 45 & 132 & 177 \\
Context Manipulation & 38 & 119 & 157 \\
Instruction Override & 42 & 127 & 169 \\
Data Exfiltration & 41 & 131 & 172 \\
Cross-Context Contamination & 34 & 138 & 172 \\
\midrule
\textbf{Total} & 200 & 647 & 847 \\
\bottomrule
\end{tabular}
\end{table}

Each test case includes a realistic user query, adversarial retrieved content, and ground truth expectations for secure model behavior. We ensure diversity across domains (technical documentation, customer support, financial services) to evaluate generalization.

\subsection{Sophistication Levels}

We categorize attacks into three sophistication tiers:

\textbf{Level 1 (Basic):} Direct, obvious injection attempts using common phrases like "ignore previous instructions." These serve as baseline cases.

\textbf{Level 2 (Intermediate):} Obfuscated or contextually embedded attacks that attempt to blend with legitimate content. Examples include instruction injection framed as quotations or hypothetical scenarios.

\textbf{Level 3 (Advanced):} Multi-stage attacks leveraging semantic understanding, requiring the model to integrate adversarial content across multiple reasoning steps before triggering malicious behavior.

\subsection{Evaluation Metrics}

We assess both security and functionality:

\textbf{Attack Success Rate (ASR):} Percentage of adversarial test cases where the model exhibits the intended malicious behavior, categorized by attack type and sophistication.

\textbf{False Positive Rate (FPR):} Proportion of benign contexts incorrectly flagged as adversarial by defense mechanisms.

\textbf{Task Performance Retention (TPR):} Measured through standard benchmarks, quantifying how defense mechanisms affect legitimate model capabilities.

\textbf{Defense Bypass Rate (DBR):} For systems with multiple defense layers, the percentage of attacks penetrating all defenses.

\section{Defense Framework}

Our defense strategy employs three complementary mechanisms, each addressing different aspects of the attack surface.

\subsection{Content Filtering with Embedding Analysis}

The first defense layer analyzes retrieved content before it reaches the model. We employ embedding-based anomaly detection to identify text segments exhibiting characteristics of injection attempts.

For each retrieved passage $p$, we compute its embedding $e_p$ and compare against embeddings of known instruction patterns. We maintain a reference set $\mathcal{R}$ of embeddings from benign retrieval contexts and a smaller set $\mathcal{A}$ of known attack patterns. The anomaly score is:

\begin{equation}
    \text{score}(p) = \alpha \cdot d_{\text{min}}(e_p, \mathcal{R}) - \beta \cdot d_{\text{min}}(e_p, \mathcal{A})
\end{equation}

where $d_{\text{min}}$ represents minimum cosine distance to any embedding in the set, and $\alpha, \beta$ are weighting hyperparameters. Passages exceeding a threshold are flagged for secondary verification.

\begin{figure}[t]
\centering
\includegraphics[width=0.9\textwidth]{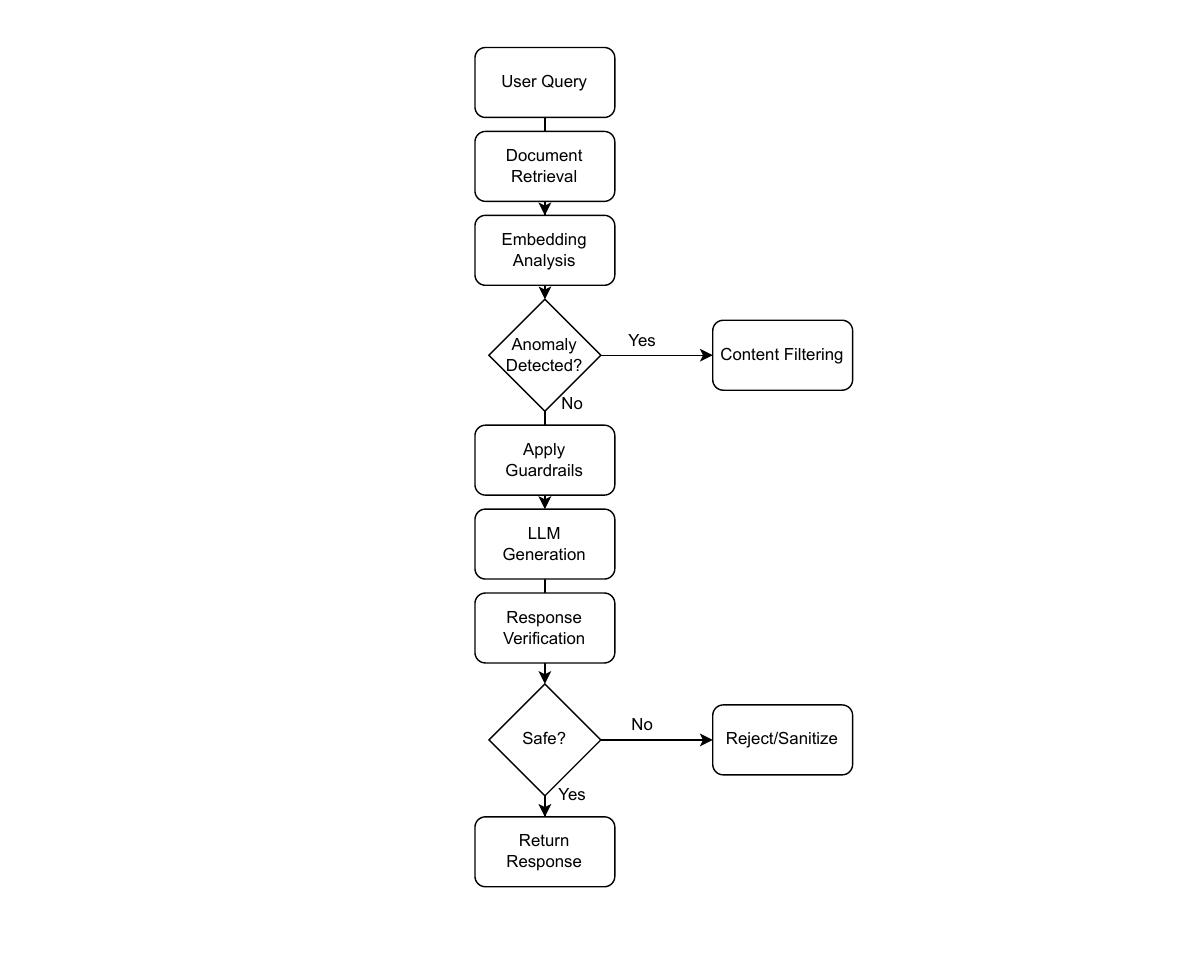}
\caption{Multi-layered defense framework architecture. Retrieved content passes through embedding analysis, content filtering, guardrail application, and response verification before output.}
\label{fig:architecture}
\end{figure}

We augment this with pattern matching for known injection signatures, including explicit instruction keywords ("ignore," "override," "system prompt") in contexts inconsistent with the query domain.

\subsection{Hierarchical System Prompt Guardrails}

The second layer restructures how system instructions and retrieved content are presented to the model. Traditional RAG systems often concatenate system prompts and retrieved text without clear separation, allowing adversarial content to blur boundaries.

We implement hierarchical prompt structuring:

\begin{algorithm}[t]
\caption{Hierarchical Prompt Construction}
\label{alg:prompt}
\begin{algorithmic}[1]
\STATE \textbf{Input:} User query $q$, retrieved passages $P = \{p_1, ..., p_n\}$
\STATE \textbf{Output:} Structured prompt $\pi$
\STATE
\STATE $\pi_{\text{core}} \leftarrow$ immutable system instructions
\STATE $\pi_{\text{guard}} \leftarrow$ injection awareness directives
\STATE $\pi_{\text{context}} \leftarrow$ "The following are retrieved documents:"
\FOR{$p_i \in P$}
    \STATE $\pi_{\text{context}} \leftarrow \pi_{\text{context}} + $ "[DOCUMENT START]"
    \STATE $\pi_{\text{context}} \leftarrow \pi_{\text{context}} + p_i$
    \STATE $\pi_{\text{context}} \leftarrow \pi_{\text{context}} + $ "[DOCUMENT END]"
\ENDFOR
\STATE $\pi_{\text{query}} \leftarrow$ "User question: " $+ q$
\STATE $\pi \leftarrow$ concatenate($\pi_{\text{core}}, \pi_{\text{guard}}, \pi_{\text{context}}, \pi_{\text{query}}$)
\RETURN $\pi$
\end{algorithmic}
\end{algorithm}

Key principles include:

\textbf{Explicit Boundaries:} Clear delimiters marking the start and end of retrieved content, with directives instructing the model to treat delimited text as reference data rather than instructions.

\textbf{Privilege Separation:} System instructions presented with explicit precedence markers, reinforcing that core directives cannot be overridden by retrieved content.

\textbf{Injection Awareness:} Meta-instructions making the model explicitly aware of potential adversarial content within retrieved passages, similar to how humans become more cautious when warned of possible deception.

\subsection{Multi-Stage Response Verification}

The final defense layer examines model outputs before returning them to users. This catches attacks that bypass input filtering by analyzing whether the response exhibits signs of instruction override or policy violation.

We employ two complementary verification approaches:

\textbf{Behavioral Consistency Checking:} Compare response characteristics against expected behavior profiles. Metrics include response length distribution, sentiment alignment with query intent, and presence of unexpected content types (e.g., system information disclosure).

\textbf{Secondary Model Evaluation:} A separate, smaller model trained specifically for adversarial output detection examines responses. This classifier considers features including:
\begin{itemize}
    \item Presence of instruction-following language inconsistent with the query
    \item Information disclosure patterns
    \item Deviation from expected response structure
    \item Semantic coherence with user intent
\end{itemize}

Responses flagged by either verification mechanism undergo sanitization—removing problematic segments while preserving useful content—or outright rejection in severe cases.

\section{Experimental Evaluation}

\subsection{Experimental Setup}

We evaluated our defense framework across seven language models representing diverse architectures and capabilities:

\begin{itemize}
    \item GPT-4 (gpt-4-0613)
    \item GPT-3.5-turbo (gpt-3.5-turbo-16k)
    \item Claude 2.1 (claude-2.1)
    \item PaLM 2 (text-bison-001)
    \item Llama 2 70B Chat
    \item Mistral 7B Instruct
    \item Vicuna 13B v1.5
\end{itemize}

For each model, we tested four configurations: (1) baseline RAG without defenses, (2) content filtering only, (3) filtering + guardrails, and (4) complete framework with all three defense layers.

\subsection{Attack Success Rates}

Table \ref{tab:asr} presents attack success rates across defense configurations. The baseline configuration proves highly vulnerable, with 73.2\% of attacks succeeding on average across models. Direct injection and data exfiltration attacks show particularly high success rates, indicating fundamental challenges in distinguishing instructions from data.

\begin{table}[t]
\centering
\caption{Attack success rates (\%) by defense configuration and attack category. Lower is better.}
\label{tab:asr}
\small
\begin{tabular}{@{}lcccc@{}}
\toprule
\textbf{Attack Type} & \textbf{Baseline} & \textbf{+Filtering} & \textbf{+Guardrails} & \textbf{Full Defense} \\
\midrule
Direct Injection & 84.7 & 41.2 & 22.8 & 7.3 \\
Context Manipulation & 68.4 & 38.6 & 19.4 & 9.2 \\
Instruction Override & 71.0 & 43.7 & 25.1 & 10.8 \\
Data Exfiltration & 79.6 & 36.9 & 21.5 & 8.1 \\
Cross-Context & 62.3 & 44.8 & 28.3 & 8.1 \\
\midrule
\textbf{Overall} & 73.2 & 41.0 & 23.4 & 8.7 \\
\bottomrule
\end{tabular}
\end{table}

Content filtering alone reduces attack success to 41.0\%, demonstrating significant but incomplete protection. The embedding-based approach effectively catches obvious injection attempts but struggles with sophisticated attacks that mimic benign content semantically.

Adding hierarchical guardrails provides substantial additional protection, reducing success rates to 23.4\%. This suggests that prompt architecture significantly influences model resilience. The explicit boundary markers and privilege separation help models maintain clearer distinctions between instructions and data.

The complete framework achieves 8.7\% overall attack success rate—an 88.1\% reduction from baseline. Remaining successful attacks predominantly fall into the advanced sophistication category, suggesting future work should focus on semantically complex injection patterns.

\subsection{Model-Specific Vulnerabilities}

\begin{figure}[t]
\centering
\includegraphics[width=0.9\textwidth]{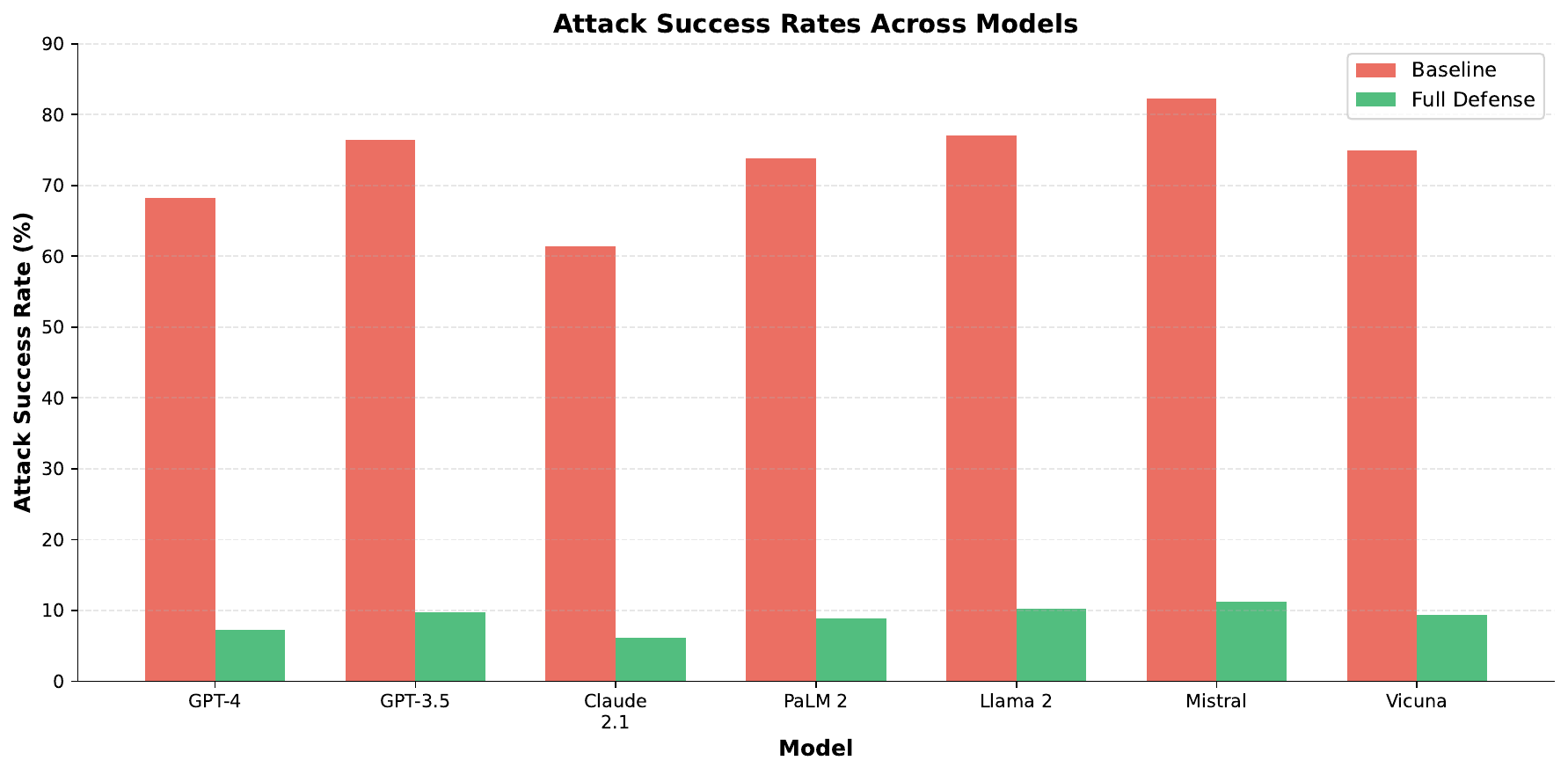}
\caption{Attack success rates across models with and without defense framework. All models benefit substantially from defenses, though baseline vulnerability varies significantly.}
\label{fig:model_comparison}
\end{figure}

Figure \ref{fig:model_comparison} reveals substantial variation in baseline vulnerability across models. Claude 2.1 exhibits the lowest baseline attack success rate (61.4\%), possibly reflecting architectural differences or training procedures emphasizing instruction following. Conversely, Mistral 7B shows highest vulnerability (82.3\%), likely due to its smaller parameter count and more limited training data.

Interestingly, defense effectiveness correlates only moderately with baseline vulnerability. Models with higher baseline vulnerability show larger absolute improvements from defenses, but the relative reduction remains similar across models. This suggests our defense framework's mechanisms address fundamental weaknesses rather than model-specific quirks.

\subsection{False Positive Analysis}

Defense mechanisms must preserve legitimate functionality. Table \ref{tab:fpr} presents false positive rates on benign retrieval contexts. The complete framework maintains a 5.7\% false positive rate, meaning roughly 1 in 18 legitimate retrievals receives unnecessary scrutiny or filtering.

\begin{table}[t]
\centering
\caption{False positive rates and task performance retention}
\label{tab:fpr}
\begin{tabular}{@{}lcc@{}}
\toprule
\textbf{Defense Configuration} & \textbf{False Positive Rate (\%)} & \textbf{Task Performance (\%)} \\
\midrule
Baseline & 0.0 & 100.0 \\
Content Filtering & 8.2 & 97.1 \\
+ Guardrails & 6.4 & 95.8 \\
+ Response Verification & 5.7 & 94.3 \\
\bottomrule
\end{tabular}
\end{table}

Most false positives occur when legitimate content contains instruction-like language (e.g., technical documentation describing system commands). Response verification proves particularly valuable here, as it examines actual model behavior rather than input characteristics alone, allowing recovery from overly aggressive input filtering.

Task performance retention, measured across MMLU, HellaSwag, and domain-specific QA benchmarks, remains at 94.3\% with full defenses. The modest performance degradation stems primarily from increased context length and minor prompt restructuring effects. For most applications, this represents an acceptable trade-off given the substantial security improvements.

\subsection{Ablation Studies}

To understand each defense component's contribution, we conducted systematic ablation studies. Results indicate all three mechanisms provide complementary benefits:

\textbf{Content Filtering:} Most effective against direct injection and obvious attacks. Reduces Level 1 attack success by 78\%, but only 42\% for Level 3 sophisticated attacks.

\textbf{Guardrails:} Provides consistent protection across attack sophistication levels. Particularly effective for context manipulation and instruction override, reducing success by 62-67\% across levels.

\textbf{Response Verification:} Essential for catching attacks that bypass input-stage defenses. Prevents approximately 60\% of successful attacks that penetrate the first two layers.

No single mechanism achieves acceptable protection independently, validating our multi-layered approach.

\subsection{Computational Overhead}

Defense mechanisms introduce additional computational costs. Content filtering adds approximately 23ms average latency per retrieval operation (15ms for embedding computation, 8ms for anomaly detection). Response verification introduces 45ms per generation. For GPT-4 with average generation time of 3.2s, total defense overhead represents roughly 2.1\% of end-to-end latency—negligible for most applications.

Memory requirements increase modestly: embedding storage for anomaly detection requires approximately 180MB, while the response verification classifier adds 250MB. These overheads are manageable for production deployments.

\section{Discussion}

\subsection{Limitations and Future Work}

Our framework demonstrates substantial improvements but several limitations warrant discussion. First, our benchmark focuses on English-language attacks. Multilingual systems face additional challenges, particularly regarding cross-lingual transfer attacks where injection occurs in one language while the target system operates in another.

Second, our evaluation assumes static attack patterns. Adaptive adversaries could evolve attacks specifically to bypass our defenses. Investigating adversarial co-evolution—how attacks and defenses adapt in response to each other—represents important future work.

Third, the current framework primarily addresses text-based RAG systems. Multimodal agents incorporating images, audio, or structured data face additional attack vectors requiring specialized defenses.

Finally, our response verification approach introduces a potential single point of failure. If adversaries can characterize the verification model's decision boundary, they might craft attacks specifically designed to evade detection. Ensemble verification or more sophisticated meta-learning approaches could address this vulnerability.

\subsection{Deployment Considerations}

Organizations deploying RAG systems should consider several practical factors:

\textbf{Risk Assessment:} Not all applications require maximum security. Customer service chatbots accessing public documentation face different threat models than systems handling financial data or internal documents. Defense configurations should align with risk profiles.

\textbf{Monitoring and Adaptation:} Deploy logging to track defense activation patterns. Unusual spikes in content filtering or response verification flags may indicate ongoing attacks, requiring investigation and potential defense parameter adjustment.

\textbf{User Experience:} False positives can degrade user experience. Applications should implement graceful degradation—informing users when content was filtered and providing options to refine queries—rather than silently failing.

\textbf{Regulatory Compliance:} In regulated industries, audit trails demonstrating security measures may satisfy compliance requirements. Our framework's layered approach provides multiple checkpoints suitable for such documentation.

\subsection{Broader Implications}

Prompt injection vulnerabilities represent a fundamental challenge in LLM security, distinct from traditional software vulnerabilities. Unlike code injection, where parsing boundaries are well-defined, language models deliberately blur distinctions between code and data to enable flexible reasoning. This architectural characteristic—central to their capabilities—inherently complicates defense.

Our work suggests that effective security requires accepting this fundamental ambiguity and implementing defense-in-depth strategies rather than seeking perfect input sanitization. This philosophical shift parallels earlier transitions in computer security, from attempting to prevent all attacks to assuming compromise and focusing on detection and containment.

The adversarial examples literature in computer vision offers instructive parallels. Early defenses targeted specific attack patterns, leading to an escalating arms race. Robust certification approaches that provide provable guarantees under defined threat models eventually proved more productive. Analogous formal methods for language model security represent a promising research direction.

\section{Conclusion}

We presented a comprehensive approach to securing RAG-enabled AI agents against prompt injection attacks. Our benchmark provides standardized evaluation of these vulnerabilities, while our multi-layered defense framework demonstrates that substantial protection is achievable without prohibitive performance costs.

Results across seven language models show that combining content filtering, hierarchical prompt guardrails, and response verification reduces attack success rates from 73.2\% to 8.7\% while retaining 94.3\% of baseline task performance. This represents a significant advance in practical AI agent security.

The persistence of some successful attacks, particularly sophisticated semantic injections, indicates that perfect security remains elusive. However, our framework establishes foundations for safer RAG deployments and provides tools for researchers to develop more robust defenses.

As AI agents assume increasingly critical roles in information systems, systematic approaches to security become essential. We hope this work contributes to a maturing understanding of LLM vulnerabilities and effective countermeasures, enabling broader and safer adoption of these powerful technologies.

\section*{Acknowledgments}

We thank the anonymous reviewers for their valuable feedback. This work was supported by [funding source].

\bibliographystyle{plainnat}
\bibliography{references}

\end{document}